# PARTICIPATING IN THE PHYSICS LAB: DOES GENDER MATTER?

BY NATASHA G. HOLMES, IDO ROLL, AND DOUGLAS A. BONN

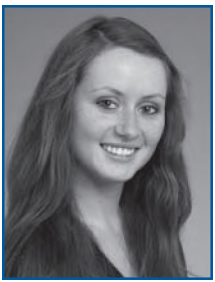

Natasha G. Holmes <nholmes@phas.ubc.ca>, Dept of Physics and Astronomy, University of British Columbia, 6224 Agricultural Road, Vancouver, BC V6T 1Z1

Ido Roll
Senior Manager, Research and Evaluation Centre for Teaching, Learning, & Technology
214-1961 East Mall, Vancouver, BC, V6T 1Z1, Canada

and

Douglas A. Bonn, Dept of Physics and Astronomy, University of British Columbia, 6224 Agricultural Road, Vancouver, BC V6T 1Z1

It is a well-studied notion that women are under-represented in the physical sciences, with a "leaky pipeline" metaphor describing how the number of women decreases at higher levels in academia [1, 2]. It is unclear, however, where the major leaks exist and what factors are responsible for this [2]. Our focus here is on women in physics with an emphasis on practical laboratory work.

A theoretical framework is under development whereby the process of learning physics (and also learning 'physicist') is described as a gendered experience. As students begin to develop an identity of what a physicist is they are also developing masculine and feminine identities of physicists [3]. The authors described how female students perceived the existence of separate male- and female-type roles in physics lab work that connect to traditional notions of femininity and masculinity. Another study found experimental evidence of this in middle school classrooms, with male students handling lab equipment significantly more often than female students during hands-on activities [4]. The group sizes in this study, however, varied between 2 and 5 students. It is possible, then, that these results are based primarily on issues of unbalanced genders in the group sizes. That is, it has previously been shown that problem solving discussions between groups with more male students than female students tended to be dominated by the male students [5]. It is, thus, not surprising that male students would also dominate with hands-on equipment if there is a gender imbalance in the group.

We aimed to study this issue further and in undergraduate classrooms through observations of how male and female students in a first-year physics lab divide roles while taking data. To address any issues of gender imbalance in the groups, we used only mixed-gender pairs of students (one male and one female student). We were testing against the null hypothesis that female students spend just as much time handling the equipment during an experiment as the male students. If the use of equipment is dominated by other psychological or sociological phenomena, then no gender effect should be observed.

## METHOD

Participants were students enrolled in a first-year honours physics course. In the lab portion of this course, students conduct an experiment each week in pairs or groups of three. The groups are randomly determined by the instructor or TAs and change each week. During the week of the study, the pairs of students were organized in a semi-random manner such that the number of mixed gender pairs (one female, one male) was maximized. Only the mixed gender pairs were included in the study. The observations took place across a single week near the end of the first term of the course when students were conducting a mass-on-a-spring experiment. The experiment asked students to determine the spring constant of a spring using Hooke's law (measuring the extension as a function of mass) and harmonic oscillation properties (measuring the period of oscillations as a function of mass).

Mixed-gender pairs were observed throughout the hands-on portion of the lab and their actions were recorded at regular time intervals as to which member of the pair was handling the equipment. One researcher discreetly monitored the class during the lab sessions and would sweep across the lab room every five minutes to record a behaviour code corresponding to each student's activity on a map of the classroom. It would take at most two minutes to sweep the whole classroom. The observer would continue to sweep until all of the students had completed their measurements and were conducting analysis or writing in their lab books. The only code included in the final analysis was which student was handling the equipment in each pair, which was then converted to a binary coding of whether the female student was using the equipment during that time interval. If neither member was actively using the equipment

SUMMARY

Observations of students in the introductory physics lab suggest that it is more common for one member of a pair to take over management of the apparatus and that gender may affect which member it is.

during an observation interval the observation was removed from the analysis (that is, the group was averaged over fewer time intervals). If both members were using the equipment during a time interval, that observation would count as half of an observation. Each pair was given a score reflecting the fraction of observations, out of those where the equipment was in use, that the female partner was the one using the equipment. That is,

$$F_{Score} = \frac{\text{\# observations female was using equipment}}{\text{\# observations equipment was being used}}. \quad (1)$$

Thus, a score of 1 means that the female was in charge of the equipment the whole time, a score of 0 means the male was in charge of the equipment the whole time, and 0.5 means they shared the usage equally.

### RESULTS

On average, the female students handled the equipment 40%±6% of the time, which was not statistically different from 50% through a one-sample t-test: $t(36) = -1.66, p = 0.106$. Fig. 1 shows the distribution of the fraction of time the female partner was using the equipment. A flat distribution would represent an equal likelihood that any individual spends any percent of time on the equipment (that is, it is just as likely that either partner would take over the equipment or that all use would be shared). A chi-square test of independence showed that the distribution in Fig. 1 may be different from a flat distribution: $\chi^2(9) = 16.24, p = 0.062$. While not significant at the 0.05 level, this result, together with the distribution, suggests that it could be more likely that one student uses the equipment the majority of the time. A moderate, positive skewness of 0.4 additionally hints that it may be the male partner who more often takes over using the equipment (demonstrated by the peak in the bin representing groups where the female only used the equipment 0–20% of the time). These results are by no means conclusive, but do motivate further investigation with a larger sample size.

### DISCUSSION

In this study, we looked at how often female students in mixed gender pairs use the equipment in a physics lab experiment compared to male students. We found evidence that male students may be more likely to take over the equipment (a large peak in the groups where the male student used the equipment more than 80% of the time). While the effect is still marginal at this point, due to a sample size of only 37 pairs, this motivates further investigation with a larger group of students. We aim to repeat the measurement this coming year to increase our sample size and explore this result further.

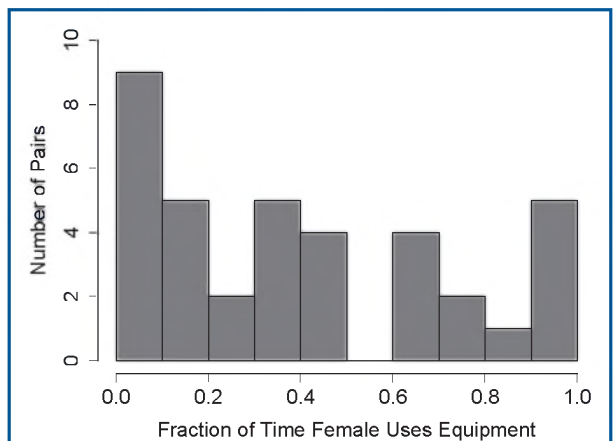

Fig. 1 The histogram shows the distribution of the fraction of observations where the female partner was using the equipment. The limits 0 and 1 represent pairs where the male or female partner was the only one using the equipment, respectively and 0.5 means the usage of equipment was split evenly between both partners.

It is likely that the use of equipment in a lab experiment is dictated by several factors such as physics knowledge, personalities, previous experience conducting experiments, and confidence levels of the group members. What this research suggests is that whichever other psychological or sociological phenomena dictate the use of lab equipment, these traits may differ by gender. Future research should examine whether any patterns of behaviours exist with same-gender pairs and include additional demographic or behavioural characteristics of students in mixed-gender pairs to identify what may be causing these differences. Future studies could also determine what classroom interventions could be used to promote female students' engagement with equipment during hands-on experiments. Any interventions, however, risk increasing students' awareness of the difference in their roles, which could further expose them to the gender stereotypes in physics, thus inducing stereotype threat[6] and reinforcing the imbalance in participation.